# Embodied negation and levels of concreteness: A TMS Study on German and Italian language processing

Giorgio Papitto[1,2], Luisa Lugli[3], Anna M. Borghi[4,5], Antonello Pellicano[6]*, and Ferdinand Binkofski[6,7]*

[1]Max Planck Institute for Human Cognitive and Brain Sciences, Department of Neuropsychology, Leipzig, Germany
[2]International Max Planck Research School on Neuroscience of Communication: Function, Structure, and Plasticity, Leipzig, Germany
[3]Department of Philosophy and Communication, University of Bologna, Bologna, Italy
[4]Department of Dynamic and Clinical Psychology, and Health Studies, Faculty of Medicine and Psychology, Sapienza University of Rome, Rome, Italy
[5]Institute of Cognitive Sciences and Technologies, National Research Council, Rome, Italy
[6]Division of Clinical Cognitive Sciences, Department of Neurology, RWTH Aachen University, Aachen, Germany
[7]Institute of Neuroscience and Medicine (INM-4), Research Center Jülich GmbH, Jülich, Germany

* These authors share senior authorship.

Correspondence should be addressed to:

Giorgio Papitto

Stephanstraße 1a, 04103 Leipzig, Germany

Phone: +49 (0) 341 9940 2238

E-mail: papitto@cbs.mpg.de

**Abstract**[1]

According to the embodied cognition perspective, linguistic negation may block the motor simulations induced by language processing. Transcranial magnetic stimulation (TMS) was applied to the left primary motor cortex (hand area) of monolingual Italian and German healthy participants during a rapid serial visual presentation of sentences from their own language. In these languages, the negative particle is located at the beginning and at the end of the sentence, respectively. The study investigated whether the interruption of the motor simulation processes, accounted for by reduced motor evoked potentials (MEPs), takes place similarly in two languages differing on the position of the negative marker. Different levels of sentence concreteness were also manipulated to investigate if negation exerts generalized effects or if it is affected by the semantic features of the sentence. Our findings indicate that negation acts as a block on motor representations, but independently from the language and words concreteness level.

---

[1] *Abbreviations*. AA, sentences with abstract verb and abstract noun; AC, sentences with abstract verb and concrete noun; Adv, adverb; ANOVA, Univariate Analyses of Variance; CC, sentences with concrete verb and concrete noun; EEG, electroencephalography; EMG, electromyography; ERD, event-related desynchronization; FDI, first dorsal interosseous; ISI, interstimulus interval; ITI, inter-trial interval; M1, primary motor cortex; MEP, motor evoked potential; MRI, Magnetic Resonance Imaging; MVPA, multi-variate pattern analysis; N, noun; pp-TMS, paired-pulse transcranial magnetic stimulation; rMT, resting motor threshold; RSVP, rapid serial visual presentation; RT, reaction time; Subj, subject; TMS, transcranial magnetic stimulation; V, verb.

# 1. Introduction

All human languages can express sentential negation through many different means that range from morphology to syntax (Dahl, 1979; De Clercq, 2020). In particular, languages that developed syntactic negation tend to differ in terms of word order, i.e., the position of the negative particle with respect to the main verb (Bernini and Ramat, 1996). For example, Italian has a Negation-Verb structure (e.g., "Io *non* ascolto"; which in English is "I don't hear"), while German has a Verb-Negation structure (e.g., "Ich höre *nicht*", semantically equivalent to the Italian example). While in Negation-Verb languages negation already introduces a representational block before the actual verbal semantics arises, the opposite happens in Verb-Negation ones, where there is a late block on semantic representations. Even if negation always contains its affirmative counterpart and shares with it its semantic content (Christensen, 2009; Greene, 1970; Kurrik, 1979), it is still under debate whether different syntactic means of negation can lead to different semantic understandings of negative sentences (Moro, 2008).

The study of negation from a psychological point of view is only of recent development. Much space has been given to its acquisition in development (Bellugi, 1967; Bloom, 1970; McNeill and McNeill, 1968; Nordmeyer and Frank, 2013; see also Dimroth, 2010), and its specific difficulty of comprehension with respect to affirmative structures (Fodor and Garrett, 1967; Gough, 1965; Haker et al., 2013; MacDonald and Just, 1989; Margolin and Abrams, 2009; Wason, 1959; Wason and Jones, 1963). However, none of these studies succeeded in tracing both a cognitive mapping of negation and a semantic understanding of how sentence polarity acts on language-derived representations. Some studies deepening these specific aspects of negation processing were provided only recently, in the context of Embodied Cognition (Alemanno et al., 2012; Bartoli et al., 2013; Beltrán et al., 2019, 2018; de Vega et al., 2016; Kaup et al., 2007b, 2006, 2005; Liuzza et al., 2011;

Lüdtke et al., 2008; Meteyard et al., 2012; Pritchett et al., 2018). According to this view, sentence comprehension is directly related to neural representations that are coherent with the action depicted by the main verb (Fischer and Zwaan, 2008). Hence, sentence comprehension is understood in terms of action simulations (Barsalou, 1999; Gallese, 2008; Glenberg and Gallese, 2012; Meteyard et al., 2012; Pezzulo et al., 2013; Taylor and Zwaan, 2008; Zwaan, 2016), even if a complete corroboration for simulations in the comprehension of abstract sentences is still weak (Dove, 2010; Rüschemeyer et al., 2007; Tettamanti et al., 2005; Tomasino and Rumiati, 2013). One line of evidence suggests that the sensorimotor system responds differently to concrete and abstract linguistic expressions both at the one-word and at the multi-word (phrase or even sentence) levels (Borghi et al., 2019, 2018, 2017; Jirak et al., 2010; Sakreida et al., 2013; Vukovic et al., 2017). Differences also emerge when combining verbs and nouns with different concreteness features, for example, a concrete verb with an abstract noun. Specifically, Scorolli et al. (2011) showed that sentences containing a concrete noun were easier to imagine than those containing an abstract noun, both when the verb preceding the noun was abstract or concrete. For what concerns negation, negative sentences have been proposed to be understood in two distinct but continuous phases: negated contents are firstly processed exactly as their affirmative counterparts, while the actual state of affairs arises only subsequently (i.e., the *two-step model hypothesis*; Kaup et al., 2007a; 2007b). However, it remains debated when these two steps occur (Liuzza et al., 2011), as well as the role of the motor cortex in sentence comprehension is unclear (Gallese and Lakoff, 2005; Paternoster, 2010; Willems and Francken, 2012). One line of evidence that bridges together concreteness and negation effects also suggests that the inhibitory effect found in negative sentences is specific only to concrete semantics (Liuzza et al., 2011). A recent multivariate pattern analysis (MVPA) study investigated sentence polarity with respect to both concrete and abstract semantics (Ghio et al., 2018). The study identifies several brain regions common to processing affirmative abstract and concrete sentences when compared to negative sentences (e.g., the left anterior and middle cingulate cortex or the precuneus). It also highlights the role of distributed representational semantic networks subserving syntactic and cognitive control systems in processing negative sentences (see also Beltrán et al., 2019). However, no direct evidence for inhibitory processes in the motor cortex and their relationship to content-specific features was provided, as well as no direct comparisons across semantic categories were performed.

A more detailed understanding of negation as a cognitive mechanism was achieved in neurophysiological experiments, where differences in polarity are the focus of investigation. Tettamanti et al. (2008) showed that processing negative action-related sentences leads to a reduction of the hemodynamic response in the frontoparietal network of the left hemisphere, often thought to be involved in action-related simulation processes (Jeannerod, 2001; Papitto et al., 2020; Pobric and Hamilton, 2006; Tettamanti et al., 2005). Such a decrease seems to be directly linked to an inhibitory process that acts on motor representations. Furthermore, Liuzza et al. (2011), in a paired-pulse transcranial magnetic stimulation (pp-TMS) experiment, found a significant modulation of motor evoked potentials (MEPs) when participants processed negative action-related sentences. However, their results do not fit in the above-described two-step model hypothesis, since this modulation occurred at an early timing, i.e., already 500 ms after sentence onset. Polarity effects were also studied through electromyography (EMG) and electroencephalography (EEG). These methodologies led to the observation that reading negative action-related relevant sentences produces a fast inhibition of the muscle congruent to the action described (Foroni and Semin, 2013) and a reduction of mu event-related desynchronization (ERD) over the motor areas (Alemanno et al., 2012). To note, these studies neither took a typological perspective nor gave a clear explanation of the time window in which negation acts. One of the studies leading to the definition of negation processing (i.e., Lüdtke et al., 2008) has been conducted on German, and the stimulus material consisted of sentences with constituent and not sentential negation (Jackendoff, 1969; Sandu, 1994). In short, constituent negation focuses its scope only on a specific element of the sentence (e.g., "Not even two years ago you could enter without paying"), while sentential negation affects the semantics of the overall sentence (e.g., "Not even two years ago could you enter without paying"; see also Haegeman, 1995). Constituent negation has been adopted experimentally also by Kaup et al. (2005), in which the negated element is a preposition of spatial relationship. In this study, the particle "not", having the function of a negative focus marker (and not of a negative polarity marker), does not deny, but it expresses a contrast and traces the path for the introduction of correct information, not acting on the tensed sentence (De Clercq, 2013; Horn, 1989). This is not the case of subsequent studies (e.g., Liuzza et al., 2011), where, instead, negation has its scope on the whole sentence and not on just one of its constituents. Furthermore, this difference does not pertain exclusively to the domains of semantics and context (Liuzza et al., 2011; Willems and Casasanto, 2011) but it also leads to strong differences in terms of syntactic structures applied (De Clercq, 2013; Horn, 1989).

The present study aims to improve knowledge on the representation of negation by addressing three main issues that have been previously overlooked: (I) typological differences in the position of the negative particle within the sentence; (II) accurate timing of inhibitory processes; and (III) interaction of negation with different sentence semantics. Concerning (I), we addressed for the first time—adopting a cross-linguistic perspective—what are the cognitive differences in processing Negation-Verb and Verb-Negation sentences, adopting languages that employ one of the two possible structures. With respect to (II), it is still unclear at what stage the motor cortex processes sentential negation. Is the inhibitory effect manifest already after the negative particle is presented? We addressed this point by employing sentences where negation had a sentential scope and by looking at its inhibitory activity within a specific time window. With regards to (III), we investigated whether only motor-related sentences are affected by the inhibitory activity of negation or whether this effect can be generalized to abstract sentences as well. To focus on this issue, we employed sentences that range from fully abstract to fully concrete: i.e., they could feature an abstract verb and an abstract noun (AA), an abstract verb and a concrete noun (AC), or a concrete verb and a concrete noun (CC).

We designed a single-pulse transcranial magnetic stimulation (TMS) experiment in which monolingual Italian (i.e., Negation-Verb language) and German (i.e., Verb-Negation language) participants had their left primary motor cortex (M1) stimulated, while they were reading sentences. Specifically, sentences were affirmative or negative at the adverb (sentential negation) and with concrete, abstract or mixed semantics that varied according to the level of concreteness associated with the verb and the noun (i.e., AA, AC, CC, as previously specified). The TMS occurred once for each sentence, 250 ms after either the onset of the verb, the noun, or the adverb (i.e., phrases).

We tested the Embodied Cognition theory for its ability to explain not only cross-cultural but also cross-linguistic phenomena (Sinha and López, 2001). Here we are specifically concerned with defining the timing in which negation is processed by the sensorimotor system. We hypothesized a reduction in the MEP signal, for both Italian and German sentences, driven by the negative markers that occurred at different sentence-specific timings (depending on the position of the adverb in each language) but at a common and word-specific timing (i.e., 250 ms after adverb presentation). This would be in line with evidence supporting that (I) semantic processing is automatically initiated immediately after the presentation of a lexical input (see also Hauk, Shtyrov, & Pulvermüller, 2008; Hinojosa, Martín-Loeches, Muñoz, Casado, & Pozo, 2004) and that (II) motor-related processes of discrimination between various semantic types (e.g., meaningful vs. meaningless, as well as motor vs. abstract lexical items) take place around 250 ms after stimulus onset (De Marco et al., 2018; Kellenbach et al., 2002; Pulvermüller et al., 2001; Scorolli et al., 2012). Moreover, we hypothesized that MEPs related to the processing of concrete sentences should be higher than those related to abstract sentences (see also Innocenti, De Stefani, Sestito, & Gentilucci, 2014; Scorolli et al., 2012). In particular, on the one hand, we expected MEPs to increase at noun position as more concrete information is introduced, thus reflecting access to more imaginable semantic features (Scorolli et al., 2011). On the other hand, at verb position, the only observable difference we expected was the one between abstract and concrete verbs, since at verb position it is not possible for participants to discriminate fully concrete sentences from sentences with mixed semantics. As such, we delivered TMS stimulations to the different phrases separately both to check for the inhibitory activity of negation at the adverb position, and to measure interactions of verbs and nouns with levels of concreteness (e.g., nouns of AA sentences vs. nouns of CC sentences).

## 2. Results

There were no significant main effects of *polarity* (affirmative, negative), *concreteness* (AA, AC, CC), and *phrase* (verb, noun, adverb; $Fs_1 < 1.158$, $ps > .32$, $\eta p^2 < .032$; $Fs_2 < 1.421$, $ps > .25$, $\eta p^2 < .066$). The main effect of *language* (Italian, German) was significant only in the item analysis ($F_1(1,36) = 1.352$, $p = .252$, $\eta p^2 = .036$; $F_2(1,20) = 314.515$, $p < .001$, $\eta p^2 = .940$).

Coherently with our hypothesis concerning a difference between affirmative and negative adverb processing, a *polarity* by *phrase* interaction resulted significant ($F_1(2, 72) = 4.188$, $p = .019$, $\eta p^2 = .104$; $F_2(2, 40) = 3.755$, $p = .032$, $\eta p^2 = .158$). Post-hoc paired sample *t*-tests revealed that MEP amplitudes at the adverb position were reduced for negative sentences (*mean₁* = 2.93 ± SD .32; *mean₂* = 2.92 ± SD .04) with respect to affirmative ones (*mean₁* = 2.95 ± SD .34, $t_1(37) = 2.618$, $p = .020$; *mean₂* = 2.94 ± SD .06, $t_2(21) = 2.595$, $p = .026$; see Figure 1). All other comparisons did not reveal significant effects, i.e. when comparing verbs in negative sentences (*mean₁* = 2.94 ± SD .34; *mean₂* = 2.94 ± SD .07) to verbs in affirmative sentences ($mean_1 = 2.94 \pm$ SD .34, $t_1(37) = -.764$, $p = .675$; $mean_2 = 2.93 \pm$ SD .07, $t_2(21) = -.702$, $p = .737$), and nouns in negative sentences ($mean_1 = 2.93 \pm$ SD .34; $mean_2 = 2.93 \pm$ SD .06) to nouns in affirmative sentences ($mean_1 = 2.93 \pm$ SD .36, $t_1(37) = -.733$, $p = .702$; $mean_2 = 2.92 \pm$ SD .06, $t_2(21) = -1.483$, $p = .230$).

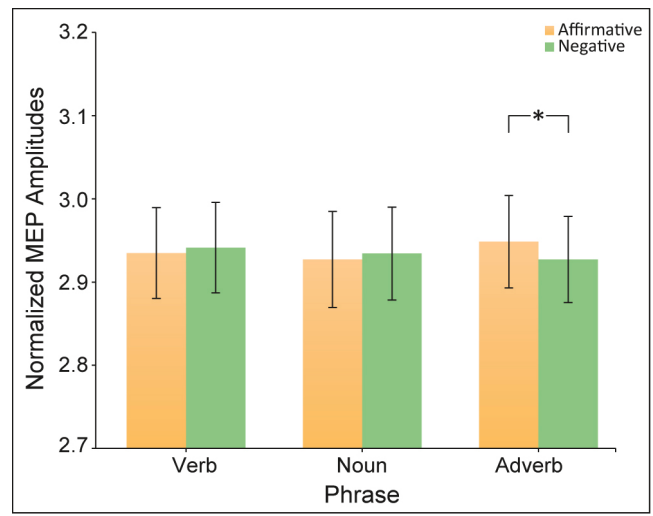

**Figure 1. MEP amplitudes relative to phrase processing for both affirmative and negative sentences.**
MEP amplitudes at Adverb in negative sentences show a significant decrease when compared to affirmative polarity sentences. Vertical bars indicate standard error means.

Concerning our expectations on a differential recruitment of the motor cortex in the concreteness spectrum, we found: (I) an interaction between *polarity* and *concreteness* for both participant and item analyses ($F_1(2,72) = 5.371$, $p = .007$, $\eta p^2 = .130$; $F_2(2, 40) = 7.001$, $p = .002$, $\eta p^2 = .259$), and (II) a *concreteness* by *phrase* interaction, which resulted significant only in the participant analysis ($F_1(4,144) = 2.499$, $p = .045$, $\eta p^2 = .065$; $F_2(2, 40) = 2.294$, $p = .066$, $\eta p^2 = .103$).

Regarding the *polarity* by *concreteness* interaction, nine post-hoc paired sample *t*-tests were performed: (I) affirmative AC vs. affirmative AA; (II) affirmative CC vs. affirmative AC; (III) affirmative CC vs. affirmative AA; (IV) negative AC vs. negative AA; (V) negative CC vs. negative AC; (VI) negative CC vs. negative AA; (VII) affirmative AA vs. negative AA; (VIII) affirmative AC vs. negative AC; and (IX) affirmative CC vs. negative CC. MEPs of AA affirmative sentences ($mean_1$ = 2.92 ± SD .35; $mean_2$ = 2.92 ± SD .06) were reduced in amplitude with respect to those of affirmative CC sentences ($mean_1$ = 2.95 ± SD .35, $t_1(37)$ = 2.944, $p$ = .027; $mean_2$ = 2.95 ± SD .06, $t_2(21)$ = 3.115, $p$ = .027; see Figure 2). Three further comparisons were significant but did not survive correction for multiple comparisons (see data analysis): (I) MEPs of AA affirmative sentences resulted smaller when compared to those of AC affirmative sentences ($mean_1$ = 2.94 ± SD .34; $t_1(37)$ = 2.405, $p$ = .095; $mean_2$ = 2.93 ± SD .07, $t_2(21)$ = 2.175, $p$ = .185); (II) MEPs of affirmative AA sentences were smaller than those of negative AA sentences ($mean_1$ = 2.94 ± SD .34; $t_1(37)$ = -2.043, $p$ = .432; $mean_2$ = 2.94 ± SD .05, $t_2(21)$ = -3.057, $p$ = .054); and (III) MEPs of affirmative CC sentences were higher than those of negative CC sentences ($mean_1$ = 2.93 ± SD .33; $t_1(37)$ = 2.032, $p$ = .442; $mean_2$ = 2.92 ± SD .05, $t_2(21)$ = 2.504, $p$ = .19).

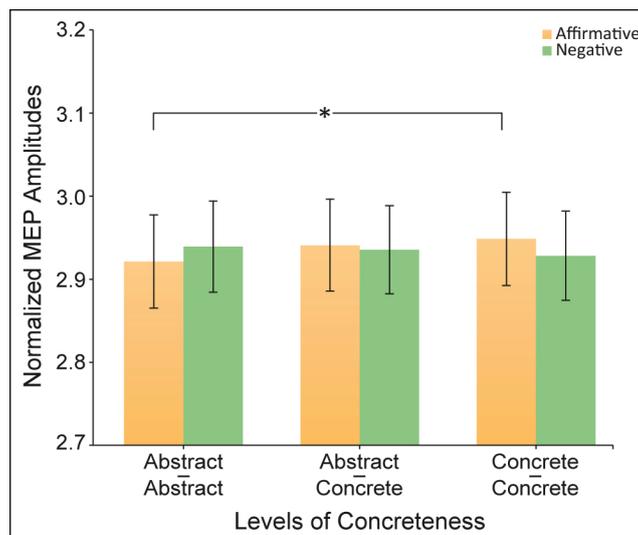

**Figure 2. MEP amplitudes of all concreteness levels for both affirmative and negative sentences.** MEP amplitudes are reduced for affirmative abstract-abstract (AA) sentences with respect to affirmative concrete-concrete (CC) sentences. Vertical bars indicate standard error means.

Concerning the *concreteness* by *phrase* interaction, six post-hoc *t*-tests were conducted: (I) AC verb vs. AA verb; (II) CC verb vs. AC verb; (III) CC verb vs. AA verb; (IV) AC noun vs. AA noun; (V) CC noun vs. AC noun; and (VI) CC noun vs. AA noun. However, these tests did not reveal any effect that survived multiple comparisons correction.

Furthermore, we found a *language* by *phrase* interaction, significant only for items as sources of variance ($F_1(2,72) = 2.178$, $p = .121$, $\eta p^2 = .057$; $F_2(2,40) = 3.760$, $p = .032$, $\eta p^2 = .158$). Three post-hoc *t*-tests were performed: (I) German verb vs. Italian verb; (II) German noun vs. Italian noun; and (III) German adverb vs. Italian adverb. All three comparisons revealed significant effects: (I) MEPs related to German verbs (*mean₂* = 2.99 ± SD .02) were higher than those related to Italian verbs (*mean₂* = 2.87 ± SD .02; $t_2(10) = 19.44$, $p < .001$); (II) the same effect was found comparing nouns in German (*mean₂* = 2.98 ± SD .02) and in Italian (*mean₂* = 2.87 ± SD .03; $t_2(10) = 9.05$, $p < .001$); (III) and it held also for adverbs, in German (*mean₂* = 2.97 ± SD .03) and in Italian (*mean₂* = 2.89 ± SD .03; $t_2(10) = 5.61$, $p < .001$).

No further interactions in both analyses resulted significant ($Fs_1 < 2.178$, $ps > .121$, $\eta p^2 < .058$; $Fs_2 < 2.467$, $ps > .132$, $\eta p^2 < .110$).

## 3. Discussion

In this study, we shed some light on negation and its early inhibitory effects, and provided further evidence for distinguishing abstract and concrete semantic features. By comparing MEPs related to processing affirmative and negative adverbs, our results suggest that the negative marker is able to inhibit the motor cortex, independently from the language under analysis. This effect occurs both when the negative marker is in a post-verbal position (e.g., "Ich schäle die Orange *nicht*"; which in English is "I do not peel the orange") or in a pre-verbal position (e.g., "Io *non* sbuccio l'arancia", semantically equivalent to the German example). We were able to isolate the effects of motor representations thanks to a rapid serial visual presentation (RSVP) method and different TMS triggering times. When a TMS stimulation occurred at the adverbial position, it was 250 ms after the onset of the negative or temporal adverb. In addition, the effect appeared not to be specific for concrete or abstract sentences. Our results point towards a more specific understanding of when negation inhibits the motor cortex. In Liuzza et al. (2011), stimulus sentences were presented without any separation between words, and a TMS stimulation was delivered randomly between 500 and 750 ms after the onset of the stimulus sentence. Given that fixation times on single words are around 300 ms (Sereno et al., 1998), TMS stimulations were delivered already when participants were involved in the lexical access of the verb. As a final remark, in Liuzza et al. (2011), MEP amplitudes related to negative Italian sentences are higher than affirmative polarity ones; thus, the effect is the opposite of what we observed. We believe that this difference can be mostly attributed to the different TMS protocols applied (see, for example, Oliveri et al., 2004). Our results also differ from those observed in a further TMS experiment on negation processing (Papeo et al., 2016). Here the authors investigated MEPs differences between affirmative and negative two-word sentences in Italian, but they were unable to detect any difference at adverb position. As a main explanation for such a discrepancy, we noted that Papeo et al. (2016) visually presented the adverb only for 250 ms and delivered the TMS stimulation 200 ms after its onset. Thus, such a stimulation time might be too early even for eliciting immediate semantic effects (Martín-Loeches et al., 2004).

Moreover, our results provide some evidence that could limit the scope of the two-step simulation hypothesis in the context of sentential negation. The observed effect of negation occurred independently from the language used and immediately after the adverb, being this before or after the verb. Hence, results provide some indication that negation acts locally since its inhibitory activity did not show any influence on other sentential phrases. According to the two-step simulation hypothesis, instead, we would have expected an effect of negation exclusively for German, where a first affirmative step could be thought as occurring before the adverb, and the second negative step occurring only at the end of the sentence, when the adverb semantics was taken into account. However, since we found an effect of negation also involving Italian, we can assume that at least in this language, the two simulations did not occur at the time of the TMS pulse, since the adverb was presented before the verb, that is, before anything could be simulated. It might also be argued that more phases are involved in processing negative sentences and what was observed in our study would only be a sub-step of a larger two-step process. Early effects (e.g., in Liuzza et al., 2011) could be dealing with negative markers as independent units, while late effects (e.g., in Lüdtke et al., 2008) might reflect processes of reanalysis where the marker is applied to the whole sentence. The absence of both the interactions language by polarity by phrase, and language by polarity by concreteness might indeed suggest that negation at this stage is not integrated into the overall sentence semantics, but it is processed as an independent element whose effects on other phrases and their concreteness level is observable only at later stages. This is a hypothesis that requires further testing and a different experimental protocol from the one used in this context. However, limiting the comparability of the two effects, many of the studies in support of the two-step simulation hypothesis employed "alternatives" (Cooper and Ginzburg, 2012). For example, in Kaup et al. (2005), participants were provided with sentences of the type "The X is (not) above/below the Y" and two images, one below the other. The task was to read a sentence and check if its content was true or not. This task is to be considered more complex not only because involving additional visual material—as stated by Bartoli et al. (2013)—but also for two further reasons. Comprehending a sentence like "The chicken is not above the egg", given the image of a chicken above an egg, could require two separate processes, exemplified as: (I) "No, the chicken is not below the egg"; (II) "The opposite is true: the chicken is above the egg". Hence, difficulty (reflected in longer reaction times; RTs) could be due to a corrective process and not to negation processing. A similar process might be required also when the negated sentence is true, that is when the same sentence is given but with the image of a chicken below the egg. Here the two processes would be: (I) "The egg is above the chicken"; (II) "Then it must be true that the chicken is not above the egg". Hence, participants focus not on the negative particle but on the spatial relationship of the items and its interaction with negation. Analogous negative constructions were used in other experiments investigating the same issue, which coherently led to an identification of response inhibition mechanisms (e.g., Clark and Chase, 1972; MacDonald and Just, 1989; Orenes et al., 2014). Similar inhibitory mechanisms were also attested in event-related potentials (ERP) studies when using negation processing in Stop-Signal (e.g., Beltrán et al., 2018) and Go/NoGo paradigms (e.g., de Vega et al., 2016). Furthermore, given the presence of two alternatives above/below, it might be argued that the negative marker "not" here does not stand for sentential negation. Indeed, we suggest that this kind of negative sentence represents a case of constituent negation. Since marked features of language tend to be more difficult to process than non-marked ones (Just and Carpenter, 1971), Kaup, Yaxley et al. (2007) observations could reflect this complexity. Conversely, in our experimental design we employed sentences where no direct alternatives were provided to the participants, thus looking directly at negation as the only process taking place. In sum, our results are not final evidence for dismissing the two-step simulation hypothesis but they question: (I) the presence of only two simulation steps, the negative one occurring exclusively after the affirmative one; and (II) the adequacy of the stimuli used for supporting this hypothesis.

The second aim of the study was to test whether the sensorimotor system reacts differently to different concreteness levels. Our stimulus material was composed of Abstract – Abstract, Abstract – Concrete, and Concrete – Concrete sentences (where the first element always refers to the verb and the second one to the noun). As expected, AA sentences showed reduced MEP amplitudes when compared to CC sentences, at the sentence level, but only for affirmative sentences. This confirms previous accounts and results showing a greater involvement of the sensorimotor system for concrete words, phrases, and sentences (Jirak et al., 2010; Klepp et al., 2019; Pérez-Gay Juárez et al., 2019; Vukovic et al., 2017). It is believed that abstract words elicit representations composed of linguistic and social information as well as emotional features, in direct contrast with words eliciting concrete representations (Borghi and Cimatti, 2009; Jirak et al., 2010). In order to account for an embodiment of abstract concepts as well, recent proposals suggest that abstract concepts are also processed similarly by the sensorimotor system (Borghi et al., 2019). However, rather than engaging the hand area of the primary motor cortex, abstract words might lead to the involvement of mouth-related areas (Borghi et al., 2011). This would account for the fact that abstract words are acquired mainly verbally through language use and social interactions (Borghi et al., 2011; Scorolli et al., 2012). Accordingly, by stimulating the hand area of the motor cortex, we found the area to be responsive to CC and not AA sentences. However, we were not able to replicate a similar result when comparing affirmative AA and AC sentences, which is abstract sentences involving concrete nouns. Given that words should be able to elicit sensorimotor representations specific to their category, we hypothesized that concrete nouns, even if in an abstract context, would elicit motor representations stronger than those of AA sentences and weaker than those of CC ones. Contrasting AA and AC sentences, an effect was observed, but it did not survive correction. It is possible that further studies directly investigating the issue might still be able to detect a difference across these levels of the concreteness continuum, but no further claims can be made in this context. Similarly, our data on the concreteness by phrase interaction are not strong enough to trace any reliable conclusion on the issue (Colquhoun, 2014). Furthermore, the distinction between AA and CC was only attested when the sentence was affirmative. No significant effects were found in the negative context. Such discrimination cannot be explained at this stage: either polarity does not interact with concreteness levels, as previously shown, or it does, as this distinction in the concreteness effect seems to suggest. Again, it is worth thinking that studies only looking at this specific distinction will lead us to a better understanding of the relationship between concreteness and polarity. Additionally, no reliable effect was observed when comparing affirmative and negative sentences across concreteness levels. This might be due to different factors. To give one speculative example, since the effect of negation is time-locked to the adverb, it is possible that MEPs related to noun and adverb processing are averaging out its inhibitory activity. This would have been cleared out by a polarity by concreteness by phrase interaction, which possibly we did not have enough power to capture. Finally, it is possible that employing a different TMS protocol could result more effective in observing small differences in the concreteness spectrum, as shown in investigating noun and verb retrieval with a pp-TMS stimulation protocol and a long interstimulus interval (ISI; Oliveri et al., 2004).

One final point to discuss concerns the discrepancy between the analysis performed with participants or items as sources of variance. For the latter, we found that independently from the phrase under analysis, MEPs related to German items were significantly greater than those of Italian items pertaining to the same syntactic class. Since we were not expecting this result, what we provide here is limited to post-hoc interpretations of the data. One reason that could lead to such a language-driven difference between items is that of the order in which the words were presented. Given the stimuli of the current experiment, German could be said to have its focus on semantic aspects of the sentence, being the Verb-Noun relationship the one that is first presented. Conversely, Italian has its focus on syntactic features of the sentence, being polarity, the first feature being processed, at the adverbial position. This difference of focus could lead to differences in the motor system's involvement in language comprehension (for a similar discussion on word-order differences, see Scorolli et al., 2011). However, the results here observed can also be linked to the possible confound for which the two groups involved in the task were entirely different (e.g., acting in different linguistic contexts).

**3.1 Limitations**

Additional considerations and limitations of this study should be discussed, especially concerning the experimental design and how it could be improved to further investigate negation processing in the brain. One important advancement would be to directly test whether interfering with the motor cortex would cause any observable effect in comprehending negative sentences. Our study did not include behavioral measures because of limits imposed by comparing the languages. Therefore, our results do not allow us to discriminate between two possible interpretations of the motor cortex's role in negative sentence processing. Indeed, further work is required to define if the concept of negation is represented in the brain as motor inhibition (strong interpretation), or if motor inhibition is only an epiphenomenon to negation processing (weak interpretation; see also Ghio & Tettamanti, 2016).

Concerning the concreteness-related effect, we were not able to further discriminate AC sentences from either AA or CC ones. This could also be related to shortcomings in the experimental design. As a matter of fact, our paradigm might not be sensitive enough to capture such distinctions in the concreteness continuum. It is possible that only focusing exclusively on this aspect—leaving other experimental manipulations out of the frame—we could learn more about how the brain shapes the boundary between sentences with both abstract and concrete elements and sentences involving only one of the two extremes of the continuum.

## 4. Conclusion

Our study indicates that the negative adverb blocks sensorimotor representations irrespectively from the language under analysis and at a specific time-point after negation. This effect is local and it occurs already 250 ms after the onset of the negative marker. Furthermore, sentences differing in levels of concreteness recruit the motor cortex differently. Affirmative CC sentences show greater MEP amplitudes when compared to fully abstract sentences, in line with previous literature on the topic. Concerning negation, our results do not support the general view of the two-step simulation hypothesis. Instead, results suggest that

negation—as soon as a lexical item introduces it in the sentential and semantic context—inhibits phrase-specific motor representations. In the context of Embodied Cognition, this entails that, to some degree, negation is processed by the sensorimotor system. The extent to which this motor contribution is crucial for negation processing as a whole is still a matter of debate. Following the evidence we provide, future research should focus on: (I) characterizing negative semantic representations in the brain, using sentences with sentential negation and including detailed behavioral measures; and (II) defining how the brain sets boundaries within the concreteness continuum and how these boundaries shape semantic processing. Addressing these two aspects separately would enhance—at a later stage—our understanding of how syntax and semantics can establish meaningful relationships of structure and content.

## 5. Experimental procedure

### 5.1 Participants

The study was approved by the ethics committee of the Medical Faculty of the RWTH Aachen University (EK 280/17). Prior to the experiment, we obtained written informed consent from 42 participants who were provided with detailed explanations about the procedure, contraindications, and risks (Rossi et al., 2009; Wassermann, 1998). We chose the number of participants taking into consideration previous TMS studies with between-participant designs (Buccino et al., 2005; Puglisi et al., 2018; Stupacher et al., 2013). Four participants were excluded from the analysis: two for technical problems with the neuronavigation system, two for difficulties in recording reliable MEPs. Of the remaining 38 participants, 19 of them were Italian native speakers (14 female, age range 20–34 years, mean age 23.1 ± 3.5 years), and 19 were German native speakers (12 female, age range 18–37 years, mean age 25 ± 4.4 years). Both Italian and German participants were recruited through flyers and outreach to RWTH Aachen University. They had neither neurological nor psychiatric diseases nor contraindications related to the single-pulse TMS procedure. Moreover, they were all right-handed, according to the Edinburgh Handedness Inventory (Oldfield, 1971), and had normal or corrected-to-normal visual acuity.

**5.2 Materials**

The stimulus material consisted of sentences, differing in terms of polarity and concreteness. For each language (German and Italian), 78 verbs were combined with 78 nouns to build 117 affirmative and 117 negative sentences. Sentences were split into three semantic groups, each one composed of 78 sentences (39 affirmatives, 39 negatives), to form a concreteness continuum: (I) Concrete verb – Concrete noun (CC), (II) Abstract verb – Concrete noun (AC), (III) Abstract verb – Abstract noun (AA; for a complete list of the stimuli, see Table S1 available at https://osf.io/gtjxp/). Of these, 18 affirmative and 18 negative sentences were selected to be used in the training session only. Examples of negative sentences were: (I) "Io non sventolo la bandiera" and "Ich schwinge die Fahne nicht" (CC; in English "I don't wave the flag"); (II) "Io non trovo la bandiera" and "Ich finde die Fahne nicht" (AC; in English "I don't find the flag"); and (III) "Io non trovo la soluzione" and "Ich finde die Lösung nicht" (AA; in English "I don't find the solution"). The same nouns were used in CC and AC sentences, as well as the same verbs in AC and AA sentences. Italian and German sentences were matched for semantic and lexical content. In CC sentences, all verbs depicted exclusively object-directed, hand-related actions, usually performed only with the dominant hand.

All verbs and nouns were extracted from the online resource of WaCky – The Web-As-Corpus Kool Yinitiative (Baroni, Bernardini, Ferraresi, & Zanchetta, 2009; available at https://wacky.sslmit.unibo.it/doku.php?id=corpora). This family of corpora contains an Italian corpus (itWac) and a German one (deWac), respectively of 1,278,177,539 and 1,585,620,279 tokens (October 2017). WaCky corpora were used not to introduce biases across languages related to different computational methodologies (Ferraresi et al., 2008). Both itWac and deWac are built applying the same web crawling procedures. To avoid any possible difference between CC, AC and AA sentences, we controlled for frequency and length of verbs and nouns between conditions. Furthermore, to avoid any kind of discrepancy between affirmative and negative sentences, in the affirmative ones, a temporal adverb was placed in the same position as the negative adverb (Papeo et al., 2016). For example, "Io *ora* sventolo la bandiera" and "Ich schwinge die Fahne *jetzt*" (in English, "I wave the flag *now*").

Univariate Analyses of Variance (ANOVA) were performed separately for verbs and nouns with word frequency (expressed in occurrences per million) and word length (expressed in number of letters) as dependent variables. Thus, four analyses were performed; all with *language* (Italian vs. German) and *concreteness* (abstract vs. concrete) as independent variables. The main effect of *language* was non-significant for noun length ($F(1,128) = 1.410$, $p = .24$) as well as for verb length ($F(1,128) = 1.103$, $p = .30$). The main effect of *concreteness* was also not significant neither for nouns ($F(1,128) = 1.193$, $p = .28$) nor for verbs ($F(1,128) = 3.187$, $p = .08$), as well as it was not significant the interaction between *language* and *concreteness* for noun length ($F(1,128) = .002$, $p = .96$) and verb length ($F(1,128) = 1.103$, $p = .30$). Similar results were obtained with regards to frequency. We found no significant main effect of *language* in both nouns ($F(1,128) = .037$, $p = .85$) and verbs ($F(1,128) = .170$, $p = .68$). Also *concreteness* resulted not significant for nouns ($F(1,128) = .894$, $p = .35$) and verbs ($F(1,128) = .195$, $p = .66$). Again, the interaction of the two variables did not show any significance for both nouns ($F(1,128) = .073$, $p = .79$) and verbs ($F(1,128) = .059$, $p = .81$; see Table 1 for the means). Finally, we checked whether AA, AC and CC sentences differed for both overall *length* and *frequency* measures. We ran two additional ANOVAs on frequency and length for nouns and verbs, with *concreteness* as independent variable. We observed no main effect of *concreteness* for both sentence length ($F(2,195) = 2.064$, $p = .130$) and frequency ($F(2,195) = .350$, $p = .705$).

Table 1

|  | Frequency | | | Length | |
| --- | --- | --- | --- | --- | --- |
|  | Noun | Verb |  | Noun | Verb |
| Italian | | | | | |
| Abstract | 43.41±48.35 | 65.32±67.54 |  | 10.55±1.67 | 6.85±1.67 |
| Concrete | 30.76±59.24 | 53.98±119.95 |  | 10.18±1.47 | 6.64±1.48 |
| Abstract and concrete | 37.08±54.03 | 59.65±96.75 |  | 10.36±1.66 | 6.74±1.56 |
| German | | | | | |
| Abstract | 42.59±58.39 | 54.47±61.79 |  | 10.9±1.98 | 7.46±1.7 |
| Concrete | 35.57±70.88 | 51.16±116.21 |  | 10.58±1.74 | 6.64±1.78 |
| Abstract and concrete | 39.08±64.53 | 52.82±92.37 |  | 10.74±1.98 | 7.05±1.78 |
| Italian and German | | | | | |
| Abstract | 42.99±53.19 | 59.90±64.46 |  | 10.72±1.91 | 7.15±1.69 |
| Concrete | 33.16±64.86 | 52.57±117.19 |  | 10.38±1.74 | 6.36±1.62 |

**Table 1. Mean values ± standard deviations for stimulus material.** Means are reported for Italian and German nouns (abstract and concrete) and verbs (abstract and concrete) for both length (numbers of letters) and frequency (occurrences per million).

### 5.3 Design and procedure

High-resolution three-dimensional T1-weighted images were first acquired on a 3 Tesla Siemens Prisma Magnetic Resonance Imaging (MRI) scanner (Erlangen, Germany). In the experimental session, muscle hot-spot was identified through neuronavigation with a frameless stereotactic system (LOCALITE Biomedical Visualization Systems GmbH, Sankt Augustin, Germany). Magnetic stimulations were delivered through a MagPro X100 stimulator equipped with an eight-shaped C-B60 coil (MagVenture A/S, Farum, Denmark) placed on the participants' head with the handle positioned in a medio-lateral direction. Visual identification of the hand area in the precentral gyrus was first performed (Yousry et al., 1997). Then, explorative stimulations were delivered to identify the exact spot of the hand area: for each stimulation, a corresponding MEP was investigated from electrodes placed on the first dorsal interosseous (FDI) muscle of the participants' right hand. For each participant, the individual resting motor threshold (rMT) was established, which is defined as the minimum intensity required to elicit MEPs with an amplitude of at least 50mV (peak-to-peak) in the FDI muscle in 5/10 consecutive trials (Rossini et al., 2015). Stimulation intensity was finally set at 120% of the participants' rMT. Participants were comfortably seated on a fully adjustable armchair (MagVenture), in front of a computer screen, at a viewing distance of 120 cm. Participants' head movements were minimized by using a headrest and a vacuum pillow. They were instructed to keep their hands and head still, and to be as relaxed as possible. The software E-Prime 2.0 (Psychology Software Tools Inc., Pittsburgh, USA) was used to present the stimuli, trigger TMS stimulations, and collect responses to control questions.

      In each trial, a fixation cross appeared at the center of the screen for 1000 ms followed by a blank screen, shown for 250 ms. Then, a sentence was presented as split into four segments (i.e., adverb (Adv), noun (N), subject (Subj), and verb (V)). These segments were arranged according to the specific grammar of each language (i.e., "Subj-Adv-V-N", for Italian; "Subj-V-N-Adv", for German). Each segment was shown for 300 ms and immediately substituted by the following one within a RSVP paradigm. RSVP and single-segment duration were adopted to accommodate natural reading speed rates (Cocklin et al., 1984; García et al., 2015; Gunter et al., 2003; Martín-Loeches et al., 2004; Martin and Altarriba, 2016; Rubin and Turano, 1992; Sereno et al., 1998; Sereno and Rayner, 2003; Spence and Witkowski, 2013). Finally, after the sentence was entirely presented, a blank screen was displayed for 4000 ms (Figure 3). Each sentence was presented twice during the course of the experiment. Short breaks were provided every 10 trials to allow the experimenter to save the collected MEPs and the participants to rest. Overall, for each language group, 198 affirmative and 198

negative sentences were randomly displayed. One third was CC, one third AC, and one third AA.

**Figure 3. Timeline of stimulus presentation during the experimental procedure.** The same rapid serial visual presentation (RSVP) method was used both for Italian and German sentences. First a fixation cross was displayed

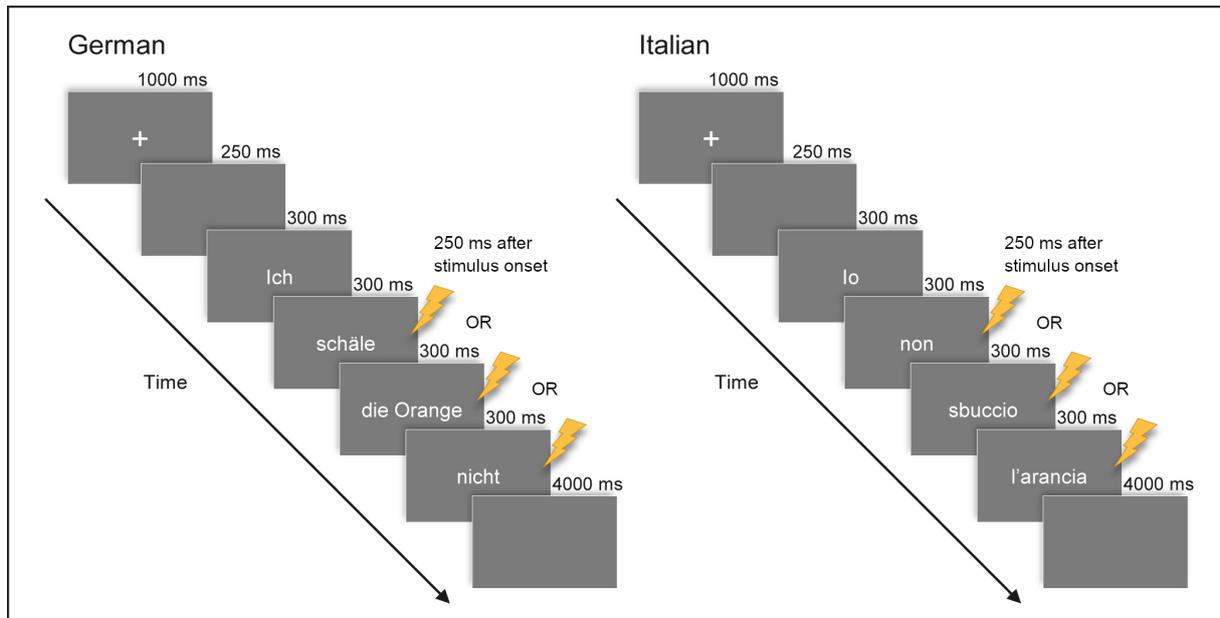

(1000 ms), followed by a blank screen (250 ms). Then, each linguistic stimulus was presented one at the time (300 ms each). The presentation always ended with a blank screen (4000 ms). Stimulation was delivered once for each sentence randomly at 250 ms after noun, verb or adverb onset.

Participants were instructed to silently read the stimulus sentences. A single-pulse TMS was pseudo-randomly delivered on the hand area of left M1 250 ms after (I) verb onset, (II) noun onset, or (III) adverb onset, with the only constraint being that an equal number of stimulations was assigned to the three different onsets. To avoid possible carry-over interference effects between TMS pulses, we decided to stimulate only one phrase per sentence, and we provided enough inter-trial interval (ITI) with the introduction of the latest slide in the RSVP, as previously described (Robertson et al., 2003; Rothwell et al., 1999). With the exclusion of the stimuli used for the training, 22 stimulations were assigned to each phrase for each concreteness level and each polarity condition. The motor cortex was stimulated in the left hemisphere given the fact that language and higher motor functions are both preliminarily processed in the left hemisphere (Aziz-Zadeh et al., 2004; Hodgson et al., 2016; Meister et al., 2006; Trettenbrein et al., 2020; van der Burght et al., 2019; Vukovic et al., 2017). The timing of the stimulation was decided on the basis of studies that revealed an early difference in the processing of concrete versus abstract words. According to this literature, action-related words are semantically processed already 220 ms after word onset (Hauk et al., 2012, 2008; Kellenbach et al., 2002; Moseley et al., 2013; Papeo et al., 2016; Scorolli et al., 2012). Furthermore, early effects of negativity in EEG studies were also attested 250 ms after onset, for non-action-related lexical items as well (Kutas and Federmeier, 2000). To check for a complete comprehension of the stimulus material, control questions were displayed at random intervals at the end of approximately 13% of the trials. Three types of questions were shown to verify that participants paid attention to the polarity of the sentence, understood the action carried out by the subject, and paid attention to the object of the action. Two possible answers were displayed on the right and the left of the screen until the participants' response was given. Participants had to respond by pressing one of two pedal buttons (7 cm x 5 cm) placed on a footboard with their right or left foot. A feedback message was displayed in case of a wrong response (for some examples of control questions and answers, see Table S2 available at https://osf.io/vtw8m/).

### 5.4 Data Analysis

All participants performed above chance in answering the control questions. The peak-to-peak amplitude (mV) of each MEP was normalized by means of a log10 logarithmic transformation (Poole et al., 2018). As a standard procedure (Candidi et al., 2010; Wilkinson et al., 2015), we discarded MEPs followed by incorrect responses to control questions (.29%), below 50mV (2.51%), and outside the range of +/- 3 SDs from the mean MEPs of each experimental condition (2.72%; for the number of trials analyzed for each experimental condition, see Table S3 available at https://osf.io/5bw7m/). Transformed amplitudes were entered in a repeated-measures ANOVA with *polarity* (affirmative, negative), *concreteness* (AA, AC, CC) and *phrase* (verb, noun, adverb) as within-participants factors. *Language* (Italian, German) was entered as a between-participants factor. Both participants and items were tested as sources of variance (Clark et al., 1973; Coleman, 1964); $F_1$ (participant analysis) and $F_2$ (item analysis) statistics were reported. In line with previous research, effects were considered reliable only when significant in both analyses (Frigo and McDonald, 1998). When necessary, paired samples *t*-tests were performed as post-hoc comparisons. Accordingly, when describing post-hoc tests, $t_1$ and *mean$_1$* refer to statistics on the participant analysis, while $t_2$ and *mean$_2$* to statistics on the item analysis. All *p* values obtained in the *t*-tests have been Bonferroni-corrected (see Bland and Altman, 1995), multiplying original values by the number of planned comparisons. Thus, only resulting values smaller than the .05 threshold were considered as significant. Where licensed by our hypotheses, we applied one-tailed *t*-tests, that is to compare (I) affirmative vs. negative phrases; and (II) AA vs. AC, AC vs. CC and AA vs. CC sentences, both affirmative and negative. In all the other cases we applied a two-tailed correction. All statistical tests were performed with the software SPSS for Windows (version 24; SPSS Inc., Chicago, IL).

**Acknowledgments**

We gratefully acknowledge H. Chen, U. S. Jawed, H. Patel, M. Marzocchi, and C. Remy for all the help provided in designing and conducting the experiment. We are also grateful to all the collaborators working at the University Hospital RWTH Aachen, where the experiment was conducted.

Funding: Giorgio Papitto was supported by the International Max Planck Research School on Neuroscience of Communication: Function, Structure, and Plasticity.

**References**

Alemanno, F., Houdayer, E., Cursi, M., Velikova, S., Tettamanti, M., Comi, G., Cappa, S.F., Leocani, L., 2012. Action-related semantic content and negation polarity modulate motor areas during sentence reading: An event-related desynchronization study. Brain Res. 1484, 39–49. https://doi.org/10.1016/j.brainres.2012.09.030

Aziz-Zadeh, L., Iacoboni, M., Zaidel, E., Wilson, S., Mazziotta, J., 2004. Left hemisphere motor facilitation in response to manual action sounds. Eur. J. Neurosci. 19, 2609–2612. https://doi.org/10.1111/j.0953-816X.2004.03348.x

Baroni, M., Bernardini, S., Ferraresi, A., Zanchetta, E., 2009. The WaCky wide web: A collection of very large linguistically processed web-crawled corpora. Lang. Resour. Eval. 43, 209–226. https://doi.org/10.1007/s10579-009-9081-4

Barsalou, L.W., 1999. Perceptual symbol systems. Behav. Brain Sci. 22, 577–660. https://doi.org/10.1017/S0140525X99002149

Bartoli, E., Tettamanti, A., Farronato, P., Caporizzo, A., Moro, A., Gatti, R., Perani, D., Tettamanti, M., 2013. The disembodiment effect of negation: Negating action-related sentences attenuates their interference on congruent upper limb movements. J. Neurophysiol. 109, 1782–1792. https://doi.org/10.1152/jn.00894.2012

Bellugi, U., 1967. The acquisition of the system of negation in children's speech. Harvard University.

Beltrán, D., Morera, Y., García-Marco, E., de Vega, M., 2019. Brain inhibitory mechanisms are involved in the processing of sentential negation, regardless of its content. Evidence from EEG theta and beta rhythms. Front. Psychol. 10, 1782. https://doi.org/10.3389/fpsyg.2019.01782

Beltrán, D., Muñetón-Ayala, M., de Vega, M., 2018. Sentential negation modulates inhibition in a stop-signal task. Evidence from behavioral and ERP data. Neuropsychologia 112,

10–18. https://doi.org/10.1016/j.neuropsychologia.2018.03.004

Bernini, G., Ramat, P., 1996. Negative Sentences in the Languages of Europe. A Typological Approach. Mouton de Gruyter, Berlin. https://doi.org/10.1515/9783110819748

Bland, J.M., Altman, D.G., 1995. Multiple significance tests: The Bonferroni method. Br. Med. J. 310, 170. https://doi.org/10.1136/bmj.310.6973.170

Bloom, L., 1970. Language development: Form and function in emerging grammars. The MIT Press, Cambridge, MA. https://doi.org/https://doi.org/10.7916/D8S75GT1

Borghi, A.M., Barca, L., Binkofski, F., Castelfranchi, C., Pezzulo, G., Tummolini, L., 2019. Words as social tools: Language, sociality and inner grounding in abstract concepts. Phys. Life Rev. 29, 120–153. https://doi.org/10.1016/j.plrev.2018.12.001

Borghi, A.M., Barca, L., Binkofski, F., Tummolini, L., 2018. Abstract concepts, language and sociality: From acquisition to inner speech. Philos. Trans. R. Soc. Lond. B. Biol. Sci. 373. https://doi.org/10.1098/rstb.2017.0134

Borghi, A.M., Binkofski, F., Castelfranchi, C., Cimatti, F., Scorolli, C., Tummolini, L., 2017. The challenge of abstract concepts. Psychol. Bull. 143, 263–292. https://doi.org/10.1037/bul0000089

Borghi, A.M., Cimatti, F., 2009. Words as tools and the problem of abstract words meanings, in: Taatgen, N., van Rijn, H. (Eds.), Proceedings of the 31st Annual Conference of the Cognitive Science Society. pp. 2304–2309.

Borghi, A.M., Flumini, A., Cimatti, F., Marocco, D., Scorolli, C., 2011. Manipulating objects and telling words: A study on concrete and abstract words acquisition. Front. Psychol. 2. https://doi.org/10.3389/fpsyg.2011.00015

Buccino, G., Riggio, L., Melli, G., Binkofski, F., Gallese, V., Rizzolatti, G., 2005. Listening to action-related sentences modulates the activity of the motor system: A combined TMS and behavioral study. Cogn. Brain Res. 24, 355–363. https://doi.org/10.1016/j.cogbrainres.2005.02.020

Candidi, M., Leone-Fernandez, B., Barber, H.A., Carreiras, M., Aglioti, S.M., 2010. Hands on the future: Facilitation of cortico-spinal hand-representation when reading the future tense of hand-related action verbs. Eur. J. Neurosci. 32, 677–683. https://doi.org/10.1111/j.1460-9568.2010.07305.x

Christensen, K.R., 2009. Negative and affirmative sentences increase activation in different areas in the brain. J. Neurolinguistics 22, 1–17. https://doi.org/10.1016/j.jneuroling.2008.05.001

Clark, H.H., Carpenter, P.A., Just, M.A., 1973. On the meeting of semantics and perception., in: Chase WG (Ed.), Visual Information Processing. Academic Press, New York, NY, pp. 311–381.

Clark, H.H., Chase, W.G., 1972. On the process of comparing sentences against pictures. Cogn. Psychol. 3, 472–517. https://doi.org/10.1016/0010-0285(72)90019-9

Cocklin, T.G., Ward, N.J., Chen, H.C., Juola, J.F., 1984. Factors influencing readability of rapidly presented text segments. Mem. Cognit. 12, 431–42. https://doi.org/10.3758/BF03198304

Coleman, E.B., 1964. Generalizing to a language population. Psychol. Rep. 14, 219–226. https://doi.org/10.2466/pr0.1964.14.1.219

Colquhoun, D., 2014. An investigation of the false discovery rate and the misinterpretation of p-values. R. Soc. open Sci. 1, 140216. https://doi.org/10.1098/rsos.140216

Cooper, R., Ginzburg, J., 2012. Negative inquisitiveness and alternatives-based negation, in: Proceedings of the 18th Amsterdam Colloquium. pp. 32–41. https://doi.org/10.1007/978-3-642-31482-7_4

Dahl, Ö., 1979. Typology of sentence negation. Linguistics 17, 79–106. https://doi.org/10.1515/ling.1979.17.1-2.79

De Clercq, K., 2020. The Morphosyntax of Negative Markers. De Gruyter Mouton, Berlin. https://doi.org/10.1515/9781501513756

De Clercq, K., 2013. A unified syntax of negation. Universiteit Gent, Ghent, Belgium.

De Marco, D., De Stefani, E., Bernini, D., Gentilucci, M., 2018. The effect of motor context on semantic processing: A TMS study. Neuropsychologia 114, 243–250. https://doi.org/10.1016/j.neuropsychologia.2018.05.003

de Vega, M., Morera, Y., León, I., Beltrán, D., Casado, P., Martín-Loeches, M., 2016. Sentential negation might share neurophysiological mechanisms with action inhibition. Evidence from frontal theta rhythm. J. Neurosci. 36, 6002–6010. https://doi.org/10.1523/JNEUROSCI.3736-15.2016

Dimroth, C., 2010. The acquisition of negation, in: Horn, L.R. (Ed.), The Expression of Negation. Mouton de Gruyter, Berlin, pp. 39–73. https://doi.org/https://doi.org/10.1515/9783110219302.39

Dove, G., 2010. On the need for embodied and dis-embodied cognition. Front. Psychol. 1, 242. https://doi.org/10.3389/fpsyg.2010.00242

Ferraresi, A., Bernardini, S., Picci, G., Baroni, M., 2008. Web corpora for bilingual lexicography. A pilot study of English/French collocation extraction and translation, in: Proceedings of The International Symposium on Using Corpora in Contrastive and Translation Studies.

Fischer, M.H., Zwaan, R.A., 2008. Embodied language: A review of the role of the motor system in language comprehension. Q. J. Exp. Psychol. 61, 825–850. https://doi.org/10.1080/17470210701623605

Fodor, J.A., Garrett, M.F., 1967. Some syntactic determinants of sentential complexity. Percept. Psychophys. 2, 289–296. https://doi.org/10.3758/BF03211044

Foroni, F., Semin, G.R., 2013. Comprehension of action negation involves inhibitory simulation. Front. Hum. Neurosci. 7, 209. https://doi.org/10.3389/fnhum.2013.00209

Frigo, L., McDonald, J.L., 1998. Properties of phonological markers that affect the acquisition of gender-like subclasses. J. Mem. Lang. 39, 218–245.

https://doi.org/10.1006/jmla.1998.2569

Gallese, V., 2008. Mirror neurons and the social nature of language: The neural exploitation hypothesis. Soc. Neurosci. 3, 317–33. https://doi.org/10.1080/17470910701563608

Gallese, V., Lakoff, G., 2005. The brain's concepts: The role of the sensory-motor system in conceptual knowledge. Cogn. Neuropsychol. 22, 455–79. https://doi.org/10.1080/02643290442000310

García, O., Cieślicka, A.B., Heredia, R.R., 2015. Nonliteral language processing and methodological considerations, in: Heredia, R. R., Cieślicka, A. (Eds.), Bilingual Figurative Language Processing. Cambridge University Press, New York, NY, pp. 117–168. https://doi.org/10.1017/CBO9781139342100.009

Ghio, M., Haegert, K., Vaghi, M.M., Tettamanti, M., 2018. Sentential negation of abstract and concrete conceptual categories: A brain decoding multivariate pattern analysis study. Philos. Trans. R. Soc. Lond. B. Biol. Sci. 373. https://doi.org/10.1098/rstb.2017.0124

Ghio, M., Tettamanti, M., 2016. Grounding sentence processing in the sensory-motor system, in: Neurobiology of Language. Elsevier, pp. 647–657. https://doi.org/10.1016/B978-0-12-407794-2.00052-3

Glenberg, A.M., Gallese, V., 2012. Action-based language: A theory of language acquisition, comprehension, and production. Cortex 48, 905–922. https://doi.org/10.1016/j.cortex.2011.04.010

Gough, P.B., 1965. Grammatical transformations and speed of understanding. J. Verbal Learning Verbal Behav. 4, 107–111. https://doi.org/10.1016/S0022-5371(65)80093-7

Greene, J.M., 1970. The semantic function of negatives and passives. Br. J. Psychol. 61, 17–22. https://doi.org/10.1111/j.2044-8295.1970.tb02797.x

Gunter, T.C., Wagner, S., Friederici, A.D., 2003. Working memory and lexical ambiguity resolution as revealed by ERPs: A difficult case for activation theories. J. Cogn. Neurosci. 15, 643–57. https://doi.org/10.1162/089892903322307366

Haegeman, L., 1995. The Syntax of Negation. Cambridge University Press, Cambridge. https://doi.org/10.1017/CBO9780511519727

Haker, H., Kawohl, W., Herwig, U., Rössler, W., 2013. Mirror neuron activity during contagious yawning--an fMRI study. Brain Imaging Behav. 7, 28–34. https://doi.org/10.1007/s11682-012-9189-9

Hauk, O., Coutout, C., Holden, A., Chen, Y., 2012. The time-course of single-word reading: Evidence from fast behavioral and brain responses. Neuroimage 60, 1462–77. https://doi.org/10.1016/j.neuroimage.2012.01.061

Hauk, O., Shtyrov, Y., Pulvermüller, F., 2008. The time course of action and action-word comprehension in the human brain as revealed by neurophysiology. J. Physiol. Paris 102, 50–58. https://doi.org/10.1016/j.jphysparis.2008.03.013

Hinojosa, J.A., Martín-Loeches, M., Muñoz, F., Casado, P., Pozo, M.A., 2004. Electrophysiological evidence of automatic early semantic processing. Brain Lang. 88, 39–46. https://doi.org/10.1016/s0093-934x(03)00158-5

Hodgson, J.C., Hirst, R.J., Hudson, J.M., 2016. Hemispheric speech lateralisation in the developing brain is related to motor praxis ability. Dev. Cogn. Neurosci. 22, 9–17. https://doi.org/10.1016/j.dcn.2016.09.005

Horn, L.R., 1989. A natural history of negation. Chicago University Press, Chicago.

Innocenti, A., De Stefani, E., Sestito, M., Gentilucci, M., 2014. Understanding of action-related and abstract verbs in comparison: a behavioral and TMS study. Cogn. Process. 15, 85–92. https://doi.org/10.1007/s10339-013-0583-z

Jackendoff, R.S., 1969. An interpretive theory of negation. Found. Lang. 5, 218–241.

Jeannerod, M., 2001. Neural simulation of action: A unifying mechanism for motor cognition. Neuroimage 14, S103–S109. https://doi.org/10.1006/nimg.2001.0832

Jirak, D., Menz, M.M., Buccino, G., Borghi, A.M., Binkofski, F., 2010. Grasping language – A short story on embodiment. Conscious. Cogn. 19, 711–720.

https://doi.org/10.1016/j.concog.2010.06.020

Just, M.A., Carpenter, P.A., 1971. Comprehension of negation with quantification. J. Verbal Learning Verbal Behav. 10, 244–253. https://doi.org/10.1016/S0022-5371(71)80051-8

Kaup, B., Ludtke, J., Zwaan, R.A., 2007a. The experiential view of language comprehension: How is negation represented?, in: Schmalhofer, F.A., Perfetti, C.A. (Eds.), Higher Level Language Processes in the Brain: Inference and Comprehension Processes. Lawrence Erlbaum, Mahwah, NJ. https://doi.org/10.4324/9780203936443

Kaup, B., Lüdtke, J., Zwaan, R.A., 2006. Processing negated sentences with contradictory predicates: Is a door that is not open mentally closed? J. Pragmat. 38, 1033–1050. https://doi.org/10.1016/j.pragma.2005.09.012

Kaup, B., Lüdtke, J., Zwaan, R.A., 2005. Effects of negation, truth value, and delay on picture recognition after reading affirmative and negative sentences, in: Bara, B.G., Barsalou, L., Bucciarelli, M. (Eds.), Proceedings of the 27th Annual Conference of the Cognitive Science Society. Lawrence Erlbaum, Mahwah, NJ, pp. 1114–1119.

Kaup, B., Yaxley, R.H., Madden, C.J., Zwaan, R.A., Lüdtke, J., 2007b. Experiential simulations of negated text information. Q. J. Exp. Psychol. 60, 976–90. https://doi.org/10.1080/17470210600823512

Kellenbach, M.L., Wijers, A.A., Hovius, M., Mulder, J., Mulder, G., 2002. Neural differentiation of lexico-syntactic categories or semantic features? Event-related potential evidence for both. J. Cogn. Neurosci. 14, 561–77. https://doi.org/10.1162/08989290260045819

Klepp, A., van Dijk, H., Niccolai, V., Schnitzler, A., Biermann-Ruben, K., 2019. Action verb processing specifically modulates motor behaviour and sensorimotor neuronal oscillations. Sci. Rep. 9, 15985. https://doi.org/10.1038/s41598-019-52426-9

Kurrik, M.J., 1979. Literature and Negation.

Kutas, Federmeier, 2000. Electrophysiology reveals semantic memory use in language

comprehension. Trends Cogn. Sci. 4, 463–470. https://doi.org/10.1016/S1364-6613(00)01560-6

Liuzza, M.T., Candidi, M., Aglioti, S.M., 2011. Do not resonate with actions: Sentence polarity modulates cortico-spinal excitability during action-related sentence reading. PLoS One 6, e16855. https://doi.org/10.1371/journal.pone.0016855

Lüdtke, J., Friedrich, C.K., De Filippis, M., Kaup, B., 2008. Event-related potential correlates of negation in a sentence-picture verification paradigm. J. Cogn. Neurosci. 20, 1355–70. https://doi.org/10.1162/jocn.2008.20093

MacDonald, M.C., Just, M.A., 1989. Changes in activation levels with negation. J. Exp. Psychol. Learn. Mem. Cogn. 15, 633–42. https://doi.org/10.1037//0278-7393.15.4.633

Margolin, S.J., Abrams, L., 2009. Not May Not be Too Difficult: The Effects of Negation on Older Adults' Sentence Comprehension. Educ. Gerontol. 35, 308–322. https://doi.org/10.1080/03601270802505624

Martín-Loeches, M., Hinojosa, J.A., Casado, P., Muñoz, F., Fernández-Frías, C., 2004. Electrophysiological evidence of an early effect of sentence context in reading. Biol. Psychol. 65, 265–80. https://doi.org/10.1016/j.biopsycho.2003.07.002

Martin, J.M., Altarriba, J., 2016. Rapid serial visual presentation: Bilingual lexical and attentional processing, in: Heredia, R., Altarriba, J., Cieślicka, A. (Eds.), Methods in Bilingual Reading Comprehension Research. Springer New York, New York, NY, pp. 61–98. https://doi.org/10.1007/978-1-4939-2993-1_4

McNeill, D., McNeill, N.B., 1968. A question in semantic development: What does a child mean when he says No?, Studies in Language and Language Behavior. Michigan University Centerfor Research on Language and Language Behavior, Ann Arbor, MI.

Meister, I.G., Sparing, R., Foltys, H., Gebert, D., Huber, W., Töpper, R., Boroojerdi, B., 2006. Functional connectivity between cortical hand motor and language areas during recovery from aphasia. J. Neurol. Sci. 247, 165–168.

https://doi.org/10.1016/j.jns.2006.04.003

Meteyard, L., Cuadrado, S.R., Bahrami, B., Vigliocco, G., 2012. Coming of age: A review of embodiment and the neuroscience of semantics. Cortex 48, 788–804. https://doi.org/10.1016/j.cortex.2010.11.002

Moro, A., 2008. The boundaries of Babel. The MIT Press, Cambridge, MA. https://doi.org/10.7551/mitpress/9780262134989.001.0001

Moseley, R.L., Pulvermüller, F., Shtyrov, Y., 2013. Sensorimotor semantics on the spot: Brain activity dissociates between conceptual categories within 150 ms. Sci. Rep. 3, 1928. https://doi.org/10.1038/srep01928

Nordmeyer, A.E., Frank, M.C., 2013. Measuring the comprehension of negation in 2-to 4-year-old children, in: Knauff, M., Pauen, M., Sebanz, N., Wachsmuth, I. (Eds.), Proceedings of the 35th Annual Conference of the Cognitive Science Society. Austin, TX, pp. 3169–3174.

Oldfield, R.C., 1971. The assessment and analysis of handedness: The Edinburgh inventory. Neuropsychologia 9, 97–113. https://doi.org/10.1016/0028-3932(71)90067-4

Oliveri, M., Finocchiaro, C., Shapiro, K., Gangitano, M., Caramazza, A., Pascual-Leone, A., 2004. All talk and no action: A transcranial magnetic stimulation study of motor cortex activation during action word production. J. Cogn. Neurosci. 16, 374–81. https://doi.org/10.1162/089892904322926719

Orenes, I., Beltrán, D., Santamaría, C., 2014. How negation is understood: Evidence from the visual world paradigm. J. Mem. Lang. 74, 36–45. https://doi.org/10.1016/j.jml.2014.04.001

Papeo, L., Hochmann, J.R., Battelli, L., 2016. The default computation of negated meanings. J. Cogn. Neurosci. 28, 1980–1986. https://doi.org/10.1162/jocn_a_01016

Papitto, G., Friederici, A.D., Zaccarella, E., 2020. The topographical organization of motor processing: An ALE meta-analysis on six action domains and the relevance of Broca's

region. Neuroimage 206, 116321. https://doi.org/10.1016/j.neuroimage.2019.116321

Paternoster, A., 2010. Le teorie simulative della comprensione e l'idea di cognizione incarnata. Sist. intelligenti XXII, 131–162. https://doi.org/10.1422/31953

Pérez-Gay Juárez, F., Labrecque, D., Frak, V., 2019. Assessing language-induced motor activity through Event Related Potentials and the Grip Force Sensor, an exploratory study. Brain Cogn. 135, 103572. https://doi.org/10.1016/j.bandc.2019.05.010

Pezzulo, G., Candidi, M., Dindo, H., Barca, L., 2013. Action simulation in the human brain: Twelve questions. New Ideas Psychol. 31, 270–290. https://doi.org/10.1016/j.newideapsych.2013.01.004

Pobric, G., Hamilton, A.F. de C., 2006. Action understanding requires the left inferior frontal cortex. Curr. Biol. 16, 524–529. https://doi.org/10.1016/j.cub.2006.01.033

Poole, B.J., Mather, M., Livesey, E.J., Harris, I.M., Harris, J.A., 2018. Motor-evoked potentials reveal functional differences between dominant and non-dominant motor cortices during response preparation. Cortex 103, 1–12. https://doi.org/10.1016/j.cortex.2018.02.004

Pritchett, B.L., Hoeflin, C., Koldewyn, K., Dechter, E., Fedorenko, E., 2018. High-level language processing regions are not engaged in action observation or imitation. J. Neurophysiol. 120, 2555–2570. https://doi.org/10.1152/jn.00222.2018

Puglisi, G., Leonetti, A., Cerri, G., Borroni, P., 2018. Attention and cognitive load modulate motor resonance during action observation. Brain Cogn. 128, 7–16. https://doi.org/10.1016/j.bandc.2018.10.006

Pulvermüller, F., Härle, M., Hummel, F., 2001. Walking or Talking?: Behavioral and Neurophysiological Correlates of Action Verb Processing. Brain Lang. 78, 143–168. https://doi.org/10.1006/brln.2000.2390

Robertson, E.M., Théoret, H., Pascual-Leone, A., 2003. Studies in cognition: The problems solved and created by transcranial magnetic stimulation. J. Cogn. Neurosci. 15, 948–60.

https://doi.org/10.1162/089892903770007344

Rossi, S., Hallett, M., Rossini, P.M., Pascual-Leone, A., 2009. Safety, ethical considerations, and application guidelines for the use of transcranial magnetic stimulation in clinical practice and research. Clin. Neurophysiol. 120, 2008–2039. https://doi.org/10.1016/j.clinph.2009.08.016

Rossini, P.M., Burke, D., Chen, R., Cohen, L.G., Daskalakis, Z., Di Iorio, R., Di Lazzaro, V., Ferreri, F., Fitzgerald, P.B., George, M.S., Hallett, M., Lefaucheur, J.P., Langguth, B., Matsumoto, H., Miniussi, C., Nitsche, M.A., Pascual-Leone, A., Paulus, W., Rossi, S., Rothwell, J.C., Siebner, H.R., Ugawa, Y., Walsh, V., Ziemann, U., 2015. Non-invasive electrical and magnetic stimulation of the brain, spinal cord, roots and peripheral nerves: Basic principles and procedures for routine clinical and research application. An updated report from an I.F.C.N. Committee. Clin. Neurophysiol. 126, 1071–1107. https://doi.org/10.1016/j.clinph.2015.02.001

Rothwell, J.C., Hallett, M., Berardelli, A., Eisen, A., Rossini, P., Paulus, W., 1999. Magnetic stimulation: Motor evoked potentials. The International Federation of Clinical Neurophysiology. Electroencephalogr. Clin. Neurophysiol. Suppl. 52, 97–103.

Rubin, G.S., Turano, K., 1992. Reading without saccadic eye movements. Vision Res. 32, 895–902. https://doi.org/10.1016/0042-6989(92)90032-E

Rüschemeyer, S.-A., Brass, M., Friederici, A.D., 2007. Comprehending prehending: Neural correlates of processing verbs with motor stems. J. Cogn. Neurosci. 19, 855–65. https://doi.org/10.1162/jocn.2007.19.5.855

Sakreida, K., Scorolli, C., Menz, M.M., Heim, S., Borghi, A.M., Binkofski, F., 2013. Are abstract action words embodied? An fMRI investigation at the interface between language and motor cognition. Front. Hum. Neurosci. 7, 125. https://doi.org/10.3389/fnhum.2013.00125

Sandu, G., 1994. Some aspects of negation in English. Synthese 99, 345–360.

https://doi.org/10.1007/BF01063993

Scorolli, C., Binkofski, F., Buccino, G., Nicoletti, R., Riggio, L., Borghi, A.M., 2011. Abstract and concrete sentences, embodiment, and languages. Front. Psychol. 2, 227. https://doi.org/10.3389/fpsyg.2011.00227

Scorolli, C., Jacquet, P.O., Binkofski, F., Nicoletti, R., Tessari, A., Borghi, A.M., 2012. Abstract and concrete phrases processing differentially modulates cortico-spinal excitability. Brain Res. 1488, 60–71. https://doi.org/10.1016/j.brainres.2012.10.004

Sereno, S.C., Rayner, K., 2003. Measuring word recognition in reading: Eye movements and event-related potentials. Trends Cogn. Sci. 7, 489–93. https://doi.org/10.1016/j.tics.2003.09.010

Sereno, S.C., Rayner, K., Posner, M.I., 1998. Establishing a time-line of word recognition: Evidence from eye movements and event-related potentials. Neuroreport 9, 2195–200. https://doi.org/10.1097/00001756-199807130-00009

Sinha, C., López, K.J. de, 2001. Language, culture, and the embodiment of spatial cognition. Cogn. Linguist. 11, 17–41. https://doi.org/10.1515/cogl.2001.008

Spence, R., Witkowski, M., 2013. Rapid Serial Visual Presentation, SpringerBriefs in Computer Science. Springer London, London. https://doi.org/10.1007/978-1-4471-5085-5

Stupacher, J., Hove, M.J., Novembre, G., Schütz-Bosbach, S., Keller, P.E., 2013. Musical groove modulates motor cortex excitability: A TMS investigation. Brain Cogn. 82, 127–36. https://doi.org/10.1016/j.bandc.2013.03.003

Taylor, L.J., Zwaan, R.A., 2008. Motor resonance and linguistic focus. Q. J. Exp. Psychol. 61, 896–904. https://doi.org/10.1080/17470210701625519

Tettamanti, M., Buccino, G., Saccuman, M.C., Gallese, V., Danna, M., Scifo, P., Fazio, F., Rizzolatti, G., Cappa, S.F., Perani, D., 2005. Listening to action-related sentences activates fronto-parietal motor circuits. J. Cogn. Neurosci. 17, 273–281.

https://doi.org/10.1162/0898929053124965

Tettamanti, M., Manenti, R., Della Rosa, P.A., Falini, A., Perani, D., Cappa, S.F., Moro, A., 2008. Negation in the brain: Modulating action representations. Neuroimage 43, 358–67. https://doi.org/10.1016/j.neuroimage.2008.08.004

Tomasino, B., Rumiati, R.I., 2013. At the mercy of strategies: The role of motor representations in language understanding. Front. Psychol. 4, 27. https://doi.org/10.3389/fpsyg.2013.00027

Trettenbrein, P.C., Papitto, G., Friederici, A.D., Zaccarella, E., 2020. Functional neuroanatomy of language without speech: An ALE meta-analysis of sign language. Hum. Brain Mapp. https://doi.org/10.1002/hbm.25254

van der Burght, C.L., Goucha, T., Friederici, A.D., Kreitewolf, J., Hartwigsen, G., 2019. Intonation guides sentence processing in the left inferior frontal gyrus. Cortex 117, 122–134. https://doi.org/10.1016/j.cortex.2019.02.011

Vukovic, N., Feurra, M., Shpektor, A., Myachykov, A., Shtyrov, Y., 2017. Primary motor cortex functionally contributes to language comprehension: An online rTMS study. Neuropsychologia 96, 222–229. https://doi.org/10.1016/j.neuropsychologia.2017.01.025

Wason, P.C., 1959. The processing of positive and negative information. Q. J. Exp. Psychol. 11, 92–107. https://doi.org/10.1080/17470215908416296

Wason, P.C., Jones, S., 1963. Negatives: Denotation and connotation. Br. J. Psychol. 54, 299–307. https://doi.org/10.1111/j.2044-8295.1963.tb00885.x

Wassermann, E.M., 1998. Risk and safety of repetitive transcranial magnetic stimulation: Report and suggested guidelines from the International Workshop on the Safety of Repetitive Transcranial Magnetic Stimulation, June 5-7, 1996. Electroencephalogr. Clin. Neurophysiol. 108, 1–16.

Wilkinson, L., Steel, A., Mooshagian, E., Zimmermann, T., Keisler, A., Lewis, J.D., Wassermann, E.M., 2015. Online feedback enhances early consolidation of motor sequence learning and reverses recall deficit from transcranial stimulation of motor cortex. Cortex. 71, 134–47. https://doi.org/10.1016/j.cortex.2015.06.012

Willems, R.M., Casasanto, D., 2011. Flexibility in embodied language understanding. Front. Psychol. 2, 116. https://doi.org/10.3389/fpsyg.2011.00116

Willems, R.M., Francken, J.C., 2012. Embodied cognition: Taking the next step. Front. Psychol. 3, 582. https://doi.org/10.3389/fpsyg.2012.00582

Yousry, T.A., Schmid, U.D., Alkadhi, H., Schmidt, D., Peraud, A., Buettner, A., Winkler, P., 1997. Localization of the motor hand area to a knob on the precentral gyrus. A new landmark. Brain 120, 141–57. https://doi.org/10.1093/brain/120.1.141

Zwaan, R.A., 2016. Situation models, mental simulations, and abstract concepts in discourse comprehension. Psychon. Bull. Rev. 23, 1028–1034. https://doi.org/10.3758/s13423-015-0864-x

**Authorship contributions**

**Giorgio Papitto**: Conceptualization, Formal analysis, Investigation, Methodology, Software, Visualization, Writing - Original Draft. **Luisa Lugli**: Conceptualization, Funding acquisition, Supervision, Writing - Review & Editing. **Anna M. Borghi**: Conceptualization, Formal analysis, Methodology, Writing - Review & Editing. **Antonello Pellicano**: Conceptualization, Formal analysis, Investigation, Methodology, Software, Supervision, Validation, Writing - Review & Editing. **Ferdinand Binkofski**: Funding acquisition, Methodology, Resources, Supervision, Writing - Review & Editing.